# Zero-bias anomaly in nano-scale hole-doped Mott insulators on a triangular silicon surface

Fangfei Ming,[1] Tyler S. Smith,[1] Steven Johnston,[1] Paul C. Snijders,[2,1] and Hanno H. Weitering[1]

[1]*Department of Physics and Astronomy, The University of Tennessee, Knoxville, Tennessee 37996*

[2]*Materials Science and Technology Division, Oak Ridge National Laboratory, Oak Ridge, Tennessee 37831*

**Abstract**

Adsorption of 1/3 monolayer of Sn on a heavily-doped p-type Si(111) substrate results in the formation of a hole-doped Mott insulator, with electronic properties that are remarkably similar to those of the high-$T_c$ copper oxide compounds. In this work, we show that the maximum hole-density of this system increases with decreasing domain size as the area of the Mott insulating domains approaches the nanoscale regime. Concomitantly, scanning tunneling spectroscopy data at 4.4 K reveal an increasingly prominent zero bias anomaly (ZBA). We consider two different scenarios as potential mechanisms for this ZBA: chiral $d_{x^2-y^2} + \mathrm{i}d_{xy}$ wave superconductivity and a dynamical Coulomb blockade (DCB) effect. The latter arises due to the formation of a resistive depletion layer between the nano-domains and the substrate. Both models fit the tunneling spectra with weaker ZBAs, while the DCB model clearly fits better to spectra recorded at higher temperatures or from the smallest domains with the strongest ZBA. Consistently, STS spectra from the lightly-doped substrates display oscillatory behavior that can be attributed to conventional Coulomb staircase behavior, which becomes stronger for smaller sized domains. We conclude that the ZBA is predominantly due to a DCB effect, while a superconducting instability is absent or a minor contributing factor.

# I Introduction

Materials exhibiting strong electronic correlations show many intriguing physical phenomena that are at the center of condensed matter physics research. For example, while the foundations of unconventional superconductivity continue to be debated [1], electron correlations appear to be playing a critical role. In high-temperature superconducting cuprates with a square lattice, the superconducting order parameter is predominately of a $d_{x^2-y^2}$ wave symmetry; however, there are reports of a subdominant $s$ or $d_{xy}$ wave symmetry component that would make the Fermi surface fully gapped [2, 3]. On the other hand, a honeycomb or triangular lattice could stabilize superconductivity with an order parameter having a spin-singlet chiral $d_{x^2-y^2} \pm \mathrm{i}d_{xy}$ wave ($d + \mathrm{i}d$) symmetry [4,5]. The nontrivial topology of the $d + \mathrm{i}d$ superconducting order parameter may give rise to Majorana modes with potential applications in quantum computing [6]. It is therefore of great interest to investigate correlated triangular or honeycomb lattice systems for possible chiral superconductivity [4,7-9].

The 'α-phases' formed by 1/3 monolayer (ML) of Sn or Pb adatoms adsorbed on the Si(111) or Ge(111) surface exhibit a (√3×√3)R30° surface reconstruction with a triangular lattice symmetry, as shown in Fig. 1. Such triangular lattices of half-filled dangling bond orbitals form a conceptually simple platform to explore two-dimensional correlated electron physics [10-16]. The α-phase of Sn on Si(111) (henceforth √3-Sn) has drawn special attention since it is a Mott insulator3Sdue to its relatively strong on-site Coulomb repulsion, $U \cong 0.6$ eV, that is of the order of the electronic bandwidth $W$ [17-20]. Our recent work has shown that the √3-Sn phase can be modulation-doped with holes when a p-type Si(111) substrate is used, reaching a maximum doping level of up to $\sim 10\%$ [21,22]. This hole-doped phase becomes metallic with a dispersing quasiparticle band that crosses the Fermi level. STS and quasiparticle interference experiments reveal that the hole-doped system has a Van Hove singularity (vHs) at 7 mV below the Fermi level and a nested constant energy contour only 10 mV above the Fermi level, suggesting that the system could be on the brink of a Fermi surface instability. Theoretical work predicts that this hole-doped system could become a $d$-wave superconductor [20] and a recent dynamical mean field theory calculation predicted a chiral $d + \mathrm{i}d$ wave superconducting phase at $> 20\%$ hole doping [23].

In this paper, we present an extensive scanning tunneling spectroscopy (STS) study on the hole-doped √3-Sn phase at low temperatures. On the most heavily doped surface, the √3-Sn phase forms isolated nano-domains surrounded by a semiconducting Si(111)(2√3×2√3)R30° surface reconstruction [22]. Our data indicate that the doping level of √3-Sn phase is higher for the smaller domains and near the edges of the √3-Sn domains; specifically, the doping level increases from $\sim 9\%$ in large domains to $\sim 12\%$ in the (much) smaller domains. Moreover, STS spectra recorded in relatively small domains show a prominent suppression of the zero bias conductance (a zero bias anomaly, ZBA for short) that becomes larger as the domain size decreases. These spectra can be fitted reasonably well with a model based on chiral $d+id$ wave superconductivity, until the size of the domain becomes too small. We also consider a dynamical Coulomb blockade (DCB) effect, specifically to model the size dependence of the ZBA [24-30]. The DCB model appears to better explain the observed spectral features, especially for the smaller nano-domains that exhibit the strongest ZBA, as well as the temperature dependence of the ZBA. Based on an analysis of an extensive data set covering variations in doping, domain size, and temperature, we conclude that the ZBA is predominantly a DCB effect. However, a scenario including superconductivity cannot be ruled out for the weaker ZBAs found on larger domains. Further experiments, possibly with higher hole-doping levels, are needed to verify the potential existence of superconductivity in these systems.

This paper is organized as follows. Sec. II outlines the experimental procedures. Sec. III presents scanning tunneling microscopy and spectroscopy (STM/STS) data about the formation of the nano-domains and determination of the hole-doping level for the different sized domains. Sec. IV presents the domain-size dependent ZBA observed at low temperature, which is then modeled in Sec. V using the chiral superconductivity and DCB scenarios mentioned above. Sec. VI presents STS data from substrates with lower carrier density, revealing classical Coulomb blockade behavior. Summary and conclusions will be presented in Sec. VII.

## II Experiment

All experiments were conducted in ultrahigh vacuum. Clean Si(111) single crystal surfaces were prepared by flash annealing up to 1200 °C in ultrahigh vacuum, followed by a short anneal at 900 °C before cooling the sample back to room temperature. Three p-type (boron doped) Si(111) substrates with room temperature resistivities of 0.001, 0.004 and 0.008 Ω·cm were used. They are labeled as B-√3, p-0.004 and p-0.008, respectively. The sample with the highest boron doping level is the B-√3 sample, which contains 1/3 ML of segregated boron atoms that are located at the $S_5$ lattice location below the surface, forming a (√3×√3)R30° superstructure [31,32].

The √3-Sn reconstruction was prepared by depositing 0.5 to 1 monolayer (ML) of Sn onto the Si(111) substrate at a substrate temperature of ~550 °C, followed by several minutes of post annealing at the same temperature. Although the √3-Sn phase has a Sn coverage of 1/3 ML, the area fraction of the √3-Sn phase on the B-√3 substrate can be maximized when more Sn is deposited, as there are competing phases on the surface [21, 22]. Reducing the total Sn coverage does not increase the area fraction of the √3-Sn phase. Instead a defective √3-Sn phase begins to dominate [21]. For the B-√3 substrate, the maximum achievable √3-Sn area fraction is ~10 % when the average Sn coverage on the surface is ~0.9 ML; for the p-0.004 and p-0.008 substrates, the maximum √3-Sn area fractions are 25 % and 85 % when ~0.6 ML and ~0.4 ML is deposited on the surface, respectively.

The samples were characterized *in situ* with an Omicron low temperature STM. A tungsten tip was used in the experiment and its metallicity was checked on an Au film before data acquisition. We obtained differential conductance ($dI/dV$) signals by superimposing a $0.6 - 5$ mV AC ripple (at 831 Hz) onto the DC tunneling voltage, and detecting the first harmonic tunneling signal with a lock-in amplifier.

## III Formation and doping level of the √3-Sn nano-domains on the B-√3 substrate

As shown in Fig. 2, the √3-Sn phase on B√3 substrates coexists with other phases. These include a high Sn-coverage phase with a (2√3×2√3)R30° reconstruction (2√3-Sn, with ~ 1.1 ML Sn coverage [22,33]) that appears higher than the √3-Sn phase in the STM topography, and amorphous islands (labeled as "AI"). The √3-Sn areas usually form at step edges but occasionally also form in the middle of terraces. These are mostly single domains without any interior domain boundaries, although domain boundaries do exist occasionally. As the neighboring 2√3-Sn phase is insulating [22], the embedded √3-Sn domains can be considered as isolated metallic islands on the semiconducting surface. Typically, √3-Sn domain sizes between approximately 200 nm$^2$ and 10000 nm$^2$ can be found on these samples, which will henceforth be designated as "small" to "large" domains, respectively.

STS dI/dV curves of the √3-Sn phase reveal the presence of three distinct spectral features within the -0.6 V to +0.6 V range, which are the hallmark of a doped single-band Mott insulator [21]. They include the upper Hubbard band (UHB), the lower Hubbard band (LHB), and the quasiparticle peak (QPP), as labeled in Fig. 3(a). The QPP peak straddles the Fermi energy at zero bias, resulting in a metallic surface. The threebroad hump-like features do not vary significantly in STS spectra taken at different temperatures (< 200 K), different locations within the same √3-Sn domain, or on different √3-Sn domains (including those of different sizes and shapes), indicating that their overall electronic structures are similar. The spectral weight of the QPP peak relative to the UHB and LHB is a measure for the doping level of the √3-Sn surface [21].

Figure 3(a) compares STS spectra recorded at 77 K along a line from the edge of a domain to its center, which is located outside the image. Evidently, the QPP intensity (or equivalently the doping level) is stronger near the edge of the domain. This larger hole doping near the edge could be due to the formation of a non-uniform space charge layer near the edge of a small metal-semiconductor contact [34], lateral charge transfer between neighboring √3 and 2√3 domains, or through doping by edge states. Such enhancement of the doping level near the edge is likely to increase the average concentration of holes more significantly for smaller sized domains. Fig. 3(b) compares the STS taken in the middle of three differently sized √3-Sn domains. The QPP on the 1230 nm$^2$ domain looks almost identical to the one obtained on the 16000 nm$^2$ large domain, whereas it apparently becomes enhanced on the small 178 nm$^2$ domain, consistent with an edge doping scenario. Fitting of these spectra with six Gaussian peaks [21] gives the percentage of the QPP intensity within the total weight of the three spectral features: 18.3 % (16000 nm$^2$), 18.8 % (1230 nm$^2$) and 25.1 % (178 nm$^2$).

These numbers convert to doping levels of 9.2 %, 9.4 % and 12.6 % respectively, with a 1.5 % error margin [21]. The fitting procedures use here are the same as those used in Ref. [21] and the reader is referred there for further details.

## IV    Zero bias anomaly in STS

For the √3-Sn surface grown on a B-√3 substrate, in addition to the three spectral features observed at 77 K, we also observe a sharp peak superimposed on the QPP very close to $E_F$ when measured at 4.4 K [Fig. 4(b)]. This additional peak has been identified as a vHs corresponding to the saddle point near the M-point in the QPP band dispersion [21]. The vHs is better resolved in STS spectra recorded with a higher energy resolution, as shown in Fig. 4(c). A significant variation of the conductance at zero bias is observed for STS taken on differently sized √3-Sn domains. For large √3-Sn domains (top curve), the vHs peak appears as a single peak centered at $\sim -7$ mV with a minor suppression at zero bias. Increasing the domain size further does not significantly alter its spectral shape. On smaller sized domains, the shape of the vHs peak does not change much but a significant suppression at zero bias appears, which becomes deeper and wider for decreasing domain sizes. The half width of the ZBA is $\sim 11$ mV for the spectrum recorded on a 232 nm$^2$ domain, and the differential conductance is suppressed by more than 60% as compared with that measured on the largest domain. The magnitude of the vHs and ZBA does not significantly depend on the shape of the domains for comparably sized domains, or on the position within a single domain, although a variation of 15% in the half-width and in the zero-bias differential conductance can be observed. The ZBA becomes fully gapped for very small domain sizes, such as the 44 nm$^2$ domain, which is the smallest √3-Sn ordered domain found in our experiment. Despite the significant suppression of the conductance within the -25 mV to 30 mV bias range, the increase of the differential conductance in the range from -100 mV to -25 mV suggests that the vHs is still present.

## V    Modeling the zero bias anomaly

In the next two subsections we will consider two potential mechanisms that might give rise to the observed ZBA. The first mechanism is a chiral $d + \mathrm{i}d$ -wave superconducting instability, proposed for the doped system in recent theoretical work [23]. The second mechanism is the DCB effect, which is rooted in the charging of the isolated √3-Sn nano-domains upon current injection in the STS experiment. We note here that energy level quantization of the √3-Sn domain due to the lateral confinement can be excluded as the source of the ZBA; a simple model of two-dimensional free electrons confined to a disc-shaped potential energy well produces an energy level separation one to two orders of magnitude smaller than the observed ZBA. Indeed, the requisite peaked quantum well state spectrum cannot be observed in the STS experiments.

**(A) Scenario I: $d + \mathrm{i}d$ superconductivity.**

Within the framework of the Hubbard model, it has been proposed that doping the triangular √3-Sn lattice could produce chiral superconductivity with an order parameter exhibiting $d + id$ symmetry [20, 23]. Generally, the superconducting pairing could be between carriers on nearest neighbor (NN), next nearest neighbor (NNN) sites, or even longer-range neighboring sites. Dynamical mean-field theory calculations, however, suggest that the pairing in such a system has mostly NN contributions [23]. We therefore use a $d + id$ wave gap function with the NN pairing to fit our spectra. The fitting formulation is similar to Ref. [35].

The spectra are modeled with a BCS density of state (DOS) broadened by a convolution with a Gaussian peak:

$$\frac{dI}{dV}(E) \propto N(E) * \text{Gaussian}(E, \sigma) \quad (1)$$

Here, $\sigma$ is the standard deviation which is set to ~2 meV to account for the modulation voltage amplitude and thermal broadening. The BCS density of states $N(E)$ is given by:

$$N(E) = -\frac{1}{\pi N} \sum_k \text{Im}\left[E - \varepsilon_k + i\Gamma - \frac{|\Delta_k|^2}{E + \varepsilon_k + i\Gamma}\right]^{-1} \quad (2)$$

where $\Gamma$ is a phenomenological scattering rate that is treated as a fitting parameter; $N$ is the number of momentum points; $\Delta_k$ is the momentum dependent superconducting gap, and $\varepsilon_k$ is the energy dispersion of the √3-Sn lattice. We model the gap function with a $d + id$ wave gap function induced by NN pairing [7]:

$$\Delta_k = \Delta_0 \left[\cos(k_x a) - \cos\left(\frac{\sqrt{3}}{2}k_y a\right)\cos\left(\frac{1}{2}k_x a\right) + i\sqrt{3}\sin\left(\frac{\sqrt{3}}{2}k_y a\right)\sin\left(\frac{1}{2}k_x a\right)\right] \quad (3)$$

where $\Delta_0$ parameterizes the gap magnitude and is another fitting parameter, and $a = 6.65$ Å is the lattice parameter of the 2D triangular lattice. The hole dispersion is modeled using a tight binding parameterization of the local density approximation (LDA) band structure given in Ref. [18]:

$$\begin{aligned}\varepsilon_k^{LDA} = &-\mu - 2t_1\left[\cos(k_x a) + 2\cos\left(\frac{\sqrt{3}}{2}k_y a\right)\cos\left(\frac{1}{2}k_x a\right)\right] \\&- 2t_2\left[\cos(\sqrt{3}k_y a) + 2\cos\left(\frac{3}{2}k_x a\right)\cos\left(\frac{\sqrt{3}}{2}k_y a\right)\right] \\&- 2t_3\left[\cos(2k_x a) + 2\cos(\sqrt{3}k_y a)\cos(k_x a)\right] \\&- 2t_4\left[\cos\left(\frac{5}{2}k_x a\right)\cos\left(\frac{\sqrt{3}}{2}k_y a\right) + \cos(2k_x a)\cos(\sqrt{3}k_y a)\right. \\&\left.+ \cos\left(\frac{1}{2}k_x a\right)\cos\left(\frac{3\sqrt{3}}{2}k_y a\right)\right] \\&- 2t_5\left[\cos(2\sqrt{3}k_y a) + 2\cos(3k_x a)\cos(\sqrt{3}k_y a)\right]\end{aligned} \quad (4)$$

where $t_1 = 52.7$ meV is the nearest neighbor hopping integral, the longer-range order hopping integrals with respect to the NN hopping are $t_2/t_1 = 0.3881$, $t_3/t_1 = 0.1444$, $t_4/t_1 = 0.0228$, and $t_5/t_1 = 0.0318$, respectively, and $\mu$ is the chemical potential, which is equal to zero for the half-filled system. A direct calculation of the DOS from the above $\varepsilon_k^{LDA}$ places the vHs at $-12$ meV, i.e. below the experimental value of $\sim -7$ mV. We therefore shift the chemical potential (Eq. 4) by $-5$ meV to account for this difference.

By placing the fitting parameters of $\Delta_0$ and $\Gamma$ into the above formalism, a simulated STS spectrum $dI/dV_{\text{simu}}(E)$ can be obtained. To compare this simulated spectrum with the experimental data, $dI/dV_{\text{simu}}(E)$ is supplemented with a linear background to account for a minor imbalance between the filled state and empty state amplitude as compared with the experimental curves. The corrected simulated spectrum is then given by:

$$dI/dV_{\text{simu}}^{\text{corr}}(E) = (1 + \alpha \cdot E) dI/dV_{\text{simu}}(E) \qquad (5)$$

with $\alpha$ being a correction factor to minimize the error between the simulated curve and the experimental curve.

The least square fits are shown in Fig. 5. The top most fit is constrained to have a vanishing superconducting gap $\Delta_0 = 0$ and the apparent ZBA region between -5 mV and 5 mV is excluded from the fitting. Therefore, the fitted curve only represents the normal state DOS including the vHs. The good quality of the fit shows that the DOS can be simply approximated by STS data recorded from a large domain that exhibits only a very small ZBA. The remaining spectra in Fig. 5 are fitted without such a constraint. The fittings down to the 232 nm² curve all look reasonable with a minor underestimation of the ZBA depth. The fitting of the 44 nm² curve is clearly poor. This discrepancy can be understood by the lack of significant coherence peaks in the experimental curve as compared with the simulation. Even though the fit to the data obtained on the 44 nm² domain is poor, this does not definitively exclude the presence of superconductivity in the larger domains, as superconductivity is expected to be suppressed in domains this small [36,37]. We also note that although higher doping level in smaller domains could enable stronger pairing tendencies, resulting in a larger ZBA, there is no significant correlation between the local doping level and the magnitude of the ZBA observed near the edge of a domain.

Finally, while we present fitting results for NN $d + id$ superconductivity here, we have also attempted to fit the spectra with pure $s$-wave, $d$-wave, and $d + id$ gap functions arising from longer range pairing. None of these gap symmetries produced a significant improvement in the fits.

**(B) Scenario II: Dynamical Coulomb blockade.**

A charging effect could also induce ZBA in the STS. In the DCB model [24], the STS experiment is modeled with a circuit consisting of a double junction, as sketched in Fig. 6. The first junction is the tunneling junction (called the T-junction) between the tip and the sample, which forms an effective capacitance $C_T$ in parallel with a resistance $R_T$. The value of $C_T$ is usually on the order of 1 aF for a typical STM tunneling junction [26] while $R_T$ is determined by the tunneling parameters as $R_T = V_S/I_t$, and is typically $R_T = 1$ $G\Omega$. The second junction is the diode junction between the √3-Sn nano-domain and the Si substrate (the S-junction) [38,39]. It is modeled using another RC circuit in series with the T-junction but with an unknown junction resistance $R$ and capacitance $C$. In all our experiments regardless of the tunneling setpoints, the energies of the spectral features (e.g. the QPP and the UHB/LHB) were always located at the same energy position, indicating that the external tunneling bias voltage is completely dropped over the T-junction, or equivalently, $R \ll R_T$. A description in terms of this DCB model is valid as long as $R$ is comparable or smaller than the resistance quantum $R_Q = 25.8$ k$\Omega$ [24]. In the DCB regime, the S-junction is treated as part of the environment and the tunneling spectra are simulated using the single electron tunneling probability across the T-junction [24]. On the other hand, the classical Coulomb blockade (CB) or Coulomb staircase (CS) have much stronger S-junction resistances with $R \gg R_Q$, and the tunneling in both junctions must be considered.

In the DCB model, the single charge tunneling probability can be calculated quantum mechanically by the $P(E)$ theory, which includes the excitation of environmental modes in the tunneling process [24,30]. The forward (e.g. tip to sample) tunneling probability is given by:

$$\vec{\Gamma}(V) = \frac{1}{e^2 R_T} \iint dE\, dE'\, n_t(E) n_s(E' - eV) f(E,T) [1 - f(E' - eV, T)]\, P(E - E') \qquad (6)$$

where $V$ is the bias voltage, $f$ is the Fermi function, $T$ is the temperature, $E$ and $E'$ are the energy levels at tip and at sample, respectively, and $n_t$ and $n_s$ are the DOS of the tip and the sample, respectively. The $P(E - E')$ term is the so called $P(E)$ function that describes the possibility of losing $E - E'$ energy to the environment ($E - E' < 0$ means energy is absorbed from the environment). In the conventional elastic tunneling picture, $P(E - E')$ reduces to a delta function $P(E - E') = \delta(E - E')$ and energy is conserved within the junction during the tunneling event. In $P(E)$ theory, the $P(E)$ function can be calculated using [24]

$$P(E) = \frac{1}{2\pi\hbar} \int_{-\infty}^{+\infty} dt\, \exp[J(t) + iEt/\hbar] \qquad (7)$$

with

$$J(t) = 2 \int_0^{+\infty} \frac{d\omega}{\omega} \frac{\mathrm{Re}\, Z(\omega)}{R_Q} \frac{e^{-i\omega t} - 1}{1 - e^{-\hbar\omega/k_B T}} \qquad (8)$$

where $k_B$ is the Boltzmann constant, $Z(\omega) = [i\omega C_T + Z_{ex}^{-1}(\omega)]^{-1}$ is the frequency dependent impedance seen from the tunneling gap, with the $Z_{ex}(\omega) = [i\omega C + 1/R]^{-1}$ being the frequency dependent impedance of the environment, which is solely composed of the S-junction in this case. The total impedance can then be written as [26]:

$$Z(\omega) = [i\omega C_\Sigma + 1/R]^{-1} \qquad (9)$$

with $C_\Sigma = C_T + C$. Considering that $C_T$ is much smaller than $C$ [26], it follows that $C_\Sigma \approx C$. With this spectral form of $Z(\omega)$, $J(t)$ can be calculated analytically [40, 41], which simplifies the fitting procedures.

The reverse tunneling is given by a similar formula:

$$\overleftarrow{\Gamma}(V) = \frac{1}{e^2 R_T} \iint dE\, dE' n_t(E) n_s(E' - eV) [1 - f(E,T)] f(E' - eV, T)\, P(E' - E) \qquad (10)$$

The total tunneling current is calculated by adding the tunneling current in both directions [24]:

$$I(V) = -e\left(\vec{\Gamma}(V) - \cev{\Gamma}(V)\right) \tag{11}$$

Finally, the simulated dI/dV($V$) spectrum is obtained by differentiating the $I(V)$ curve followed by a convolution with a Gaussian line shape [see Eq. (1)] to capture instrument broadening.

Given $n_t$ and $n_s$, the tunneling spectrum based on the DCB model can be obtained by varying the $R$ and $C$ parameters of the S-junction. In our calculation, $n_t$ is taken as a constant since the tip is ensured to be metallic and qualitatively featureless. As we discussed in Fig. 5, $n_s$ is approximated by the spectrum taken on large domains, but with the minor suppression between -5 mV and 5 mV removed. Specifically, we construct $n_s$ using the smoothed experimental curve obtained from a large domain. We then replace the $n_s$ values in the [-5,5] meV range with a polynomial fitting to the surrounding data points. This procedure produces a constructed DOS with no ZBA, which is labeled as the "DOS" curve shown in Fig. 6. This constructed DOS reproduces the spectral behaviors at higher energies beyond the ZBA influenced regions, such that the fit of the DCB model to the data will be able to optimize mainly on the ZBA features. It should be noted that we have verified that different ways of obtaining the $n_s$ curve only have very minor effects on the quality of the fits; the fitting parameters mainly depend on the shape of the ZBA in the spectra.

The data set shown in Fig. 4(c) were fitted with the DCB model using adjustable parameters $R$ and $C$ of the S-junction. The capacitance $C$ mainly determines the depth and width of the ZBA while the resistance $R$ mainly determines the steepness of the ZBA. As seen in Fig. 6, the fitted curves account for the ZBA quite well, even for the wide gap observed on the 44 nm² domain. For the curves recorded in the 1215 nm², 412 nm², and 232 nm² domains, the fits begin to systematically underestimate the first peak just beyond the ZBA at positive bias, e.g. at ~10 mV.

Solely comparing the quality of the fits produced by the superconductivity model in Fig. 5 and the DCB model in Fig. 6, it appears that the superconductivity scenario fits slightly better in cases where ZBA does not produce a zero differential conductance at zero bias voltage. Such advantage would be more significant by noting that the fitting with DCB model depend on an external input of the $n_s$, whereas the superconductivity model does not. Besides that, the two models have equal number of fitting parameters, e.g., $\Delta_0$ and $\Gamma$ for the superconductivity model; $R$ and $C$ for DCB model. However, these differences are quite minor and do not really discriminate against the DCB model. On the other hand, the DCB model does a significantly better job in capturing the hard gap in the 44 nm² curve.

In the DCB model, the S-junction is formed between the √3-Sn domains and the Si substrate that are separated by an insulating depletion layer. The charging effect due to the S-junction will be enhanced at smaller sized domains, e.g. larger $R$ and smaller $C$. Assuming the junction is laterally uniform, which is a valid assumption since the depletion layer is only ~ 1 nm thick (see below) compared with the lateral sizes of the domains between 10-100 nm, the junction parameters $R$ and $C$ should scale with the domain sizes: $C = \alpha_C \cdot A$ and $R = \alpha_R \cdot A^{-1}$, where $\alpha_C$ is the capacitance per unit area, $\alpha_R$ the resistivity per reverse unit area, and $A$ is the area of the domain. Fig. 7 plots the $R$ and $C$ parameters obtained from STS taken on a series of domains with different sizes, where their linear dependence on the size or inverse size is apparent. A linear fit of the data points provides an estimate of the scaling factors: $\alpha_C = 0.081 \text{ aF/nm}^2$ and $\alpha_R = 310 \text{ R}_Q \cdot \text{nm}^2$. These values are consistent with the estimated surface doping concentration. If we take the relative permittivity of Si as 11.7 and use $C = \varepsilon_r \varepsilon_0 A/d$ ($\varepsilon_0$ is the vacuum permittivity and $d$ is the depletion width), a depletion width $d$ = 1.3 nm is obtained [36]. If we then assume that the conductance across the S-junction occurs via tunneling (which should be valid at our low temperature experiments [36]) and use a barrier height of ~ 0.7 V obtained from XPS measurements [21], the resistance per reverse unit area $\alpha_R$ converts to an effective concentration of ionized dopants of $1 \times 10^{20} \text{ cm}^{-3}$ in the depletion layer [36]. Transferring all carriers originating from these ionized charges in the depletion layer to the surface induces a ~ 5 % hole doping in the √3-Sn lattice. This value is reasonable in comparison to the 9 – 12 % hole doping level estimated from spectral weight analysis of the STS in the discussion of Fig. 4. It should be noted that this is a crude estimation; e.g., $\varepsilon_r$ increases for heavily doped Si [42], which would result in a larger estimate of the surface doping level.

**(C) Temperature dependent ZBA.**

The ZBA and the vHs also depend on temperature. Fig. 8 compares the temperature dependent STS measurement on three √3-Sn domains, which are representative of the STS measurements on different domains. Fig. 7(a) shows the STS taken on a 12370 nm² domain with no apparent ZBA at all temperatures. The black curves in Fig. 8(b) and 8(c) are the temperature dependent spectra taken on two smaller domains, with sizes of 1148 nm² (b,d) and 243 nm² respectively, showing significant ZBA at low temperatures. As the temperature rises, both the vHs and the ZBA become weaker and vanish above 20 K for the 1148 nm² data set or above 50 K for the 243 nm² data set. Temperature dependent STS measured on a 44 nm² domain persist above 77 K (4.4 K data shown in Fig. 4, higher temperature data not shown). The red curves in Fig. 8(b) and 8(c) show the fitted/simulated spectra from the DCB model, in which the large domain STS spectra in Fig. 8(a) are smoothed and serve as $n_s$ in fitting Eq. (6) and Eq. (10). Only the 4.4 K spectra is fit, giving a best fitting parameter of the S-junction as $R = 0.39\, R_Q$ and $C = 147$ aF for the 1148 nm² domain or $R = 1.5\, R_Q$ and $C = 30$ aF for the 243 nm² domain. The same junction parameters are then used to simulate higher temperature spectra. They reproduce the spectra well in the region where the ZBA forms. This consistency supports a description of the ZBA in terms of the DCB. The fact that the junction parameters do not change significantly with temperature indicates that the depletion width remains roughly constant up to 77 K, and the conductance through the junction is still mainly through tunneling.

We also attempt to fit the same set of data with the $d + id$ model. However, the significant temperature dependence of the vHs makes it difficult to fit the ZBA with the fitting formulas [Eqs. (2)-(5)]. These equations only contain an implicit temperature-dependent variable in the gap magnitude $\Delta_0$ and scattering parameter $\Gamma$ which are expected to account for the ZBA feature, but do not account for the temperature dependence of the vHs. We therefore calculate the BCS density state with the Dynes function [43,44], meaning that Eq. (2) is replaced with:

$$N_D(E,T) = \frac{1}{\pi} n_s(E,T) \int_0^\pi d\theta \cdot Re\left(\frac{E - i\Gamma}{\sqrt{(E-i\Gamma)^2 - |\Delta(\theta)|^2}}\right) \quad (11)$$

where the $\Delta(\theta)$ is the momentum direction ($\theta$) dependent gap function at Fermi wave vector $k_F(\theta)$. $\Delta(\theta)$ could be obtained by Eq. (3) and Eq. (4). $n_s(E,T)$ is the normal state DOS which could be approximated by the smoothed STS on large domain in Fig. 7(a). The $I(V)$ curve is simulated by

$$I(V) = \int dE \cdot N_D(E - eV, T)[f(E,T) - f(E - eV, T)] \quad (12)$$

Finally, the simulated $dI/dV(V)$ spectrum is obtained by differentiating the $I(V)$ curve followed by a convolution with a Gaussian line shape [see Eq. (1)] to capture instrument broadening. The fits are shown in Fig. 8(d) and 8(e). The $d + id$ model fits well to the 1148 nm² data set and the fitting quality is similar to or slightly better than the DCB model. On the other hand the fitting quality to the 243 nm² data set is clearly poorer at low temperatures as compared to the DCB model. Although we have tried different background subtraction methods or different gap functions for fitting, the quality of the fits for the small domain is consistently inferior to those based on the DCB model. Moreover, for both 1148 nm² data set and the 243 nm² data set, the fitted superconducting gaps appear to increase from the 4.4 K value before it starts to drop, which seems to be inconsistent with a superconductivity scenario.

## VI  ZBA on other substrates

Further support of the Coulomb charging effect is obtained from the same √3-Sn nano-domains grown on lesser doped substrates. Here, we analyze STS spectra recorded on √3-Sn domains grown on p-0.004 and p-0.008 substrates, which are expected to have a wider depletion layer due to their lower boron dopant concentration. Therefore, they are expected to exhibit stronger Coulomb charging effects than the B√3 substrate. On these two substrates, the area fraction of the √3-Sn phase is higher, and it is more common for the √3-Sn domains to neighbor one another. However, the maximum size of the √3-Sn domains is not significantly larger than that on the B√3 substrate due to the presence of domain boundaries, as indicated in Fig. 9(a). As will be shown in the following, these domain boundaries electronically separate neighboring domains, resulting in distinctly different STS spectra in terms of the ZBA.

Figures 9(b) and 9(c) compare the STS on large (~ 5000 nm²) and small (~250 nm²) domains on the p-0.004 and p-0.008 substrates, together with the STS from the previously discussed B√3 substrate in Fig. 4(c), all measured at 4.4 K. In Fig. 9(b), the STS spectra recorded on large domains on different substrates show very different behaviors in terms of the vHs and ZBA features. The B-√3 curve has a very strong vHs but with a weak ZBA. The vHs feature is significantly smaller but still observable on the p-0.004 sample, with a much stronger ZBA. Finally, the vHs feature for the p-0.008 is completely gone, leaving only a possible shoulder at -20 mV, while the ZBA is even stronger, almost resembling a hard gap in the spectrum. These differences could be attributed in part to changes of the QPP associated with the different doping levels of the √3-Sn phase on these substrates [21]. The enhancement of ZBA appears to be consistent with the fact that the depletion layer will be wider for the p-0.004 and p-0.008 substrate compared to the more heavily doped B-√3 substrate. There may, however, be remnants of a Mott gap in the spectra recorded on the p-0.004 and p-0.008 substrates [21], complicating an explanation of the observed differences.

For small domains on the p-0.004 and p-0.008 substrates, as well as the B-√3 substrate, as shown in Fig. 9(c), the spectral features at higher energies remain unchanged as compared with their counterparts recorded on the large domains in Fig. 9(b), while the ZBAs are all enhanced. The ZBA for the p-0.008 sample reveals a staircase-like feature, which appear as a hard gap with several plateaus in the dI/dV spectrum. The CS behavior happens when the S-junction resistance is comparable to or even larger than the T-junction resistance. This staircase feature is weaker but still observable in the "p-0.008" large domain STS [in Fig. 9(b)] as well as in the "p-0.004" small domain STS [in Fig. 9(c)], consistent with their expected weaker S-junction. These staircase features in the STS appears to be superimposed over a V-shaped background, which could be understood as the pseudogap-like feature in the STS of hole doped Mott insulators [21]. This residual Mott feature is only completely absent on the B√3 substrate, which represents the highest doping level in our experiment [21]. However, qualitatively, the systematic enhancement of the ZBA in STS from large domain to small domain clearly indicates the charging effect is universal for different substrates. The charging effect could manifest as DCB, CB, and CS features in the STS, depending on both the S-junction and the T-junction characteristics. The superconductivity picture clearly fails to explain the behavior of the ZBA on these two substrates.

It is worth noting that the neighboring large and small domains on the p-0.008 substrate, shown as "L" and "S" in Fig. 9(a), have very different behavior in the ZBA. Similar observations are made for the p-0.004 and B√3 substrates. This commonality indicates that the domain boundary of the √3-Sn phase electronically decouples neighboring domains, and the electrons tunneling into a nano-domain are therefore forced to drain through the depletion layer in the substrate. The insulating domain boundary prohibits the √3-Sn phase from forming a metallic film macroscopically, which is expected to influence the transport measurement of such surface [45].

Since the STS data of the √3-Sn nano-domains on the p-0.004 and p-0.008 substrates indicate that the system is in the classical CB or CS regime (meaning the T-junction and S-junction become comparable and both will influence the tunneling process), tuning the T-junction by changing the tip sample separation should result in a modified ZBA behavior. Fig. 10 shows such tip-sample separation dependent STS measured on a large √3-Sn domain on the p-0.008 substrate. Fig. 10(a) measures wider energy scale where the QPP and part of the UHB/LHB are resolved. With progressively reduced tip-sample separation, these spectral features all move toward higher energy, indicating a tip induced bend bending effect due to the poor screening in the Si bulk [46]. This effect should be distinguished from the Coulomb charging effect, as it is purely a static electrical field effect. Conversely, the Coulomb charging effect is a non-equilibrium dynamical process of the electron tunneling. Therefore, the two mechanisms work independently. A similar tip-sample separation dependent measurement on the B-√3 or the p-0.004 sample does not show a tip induced bend bending effect (data not shown), consistent with the better screening in these substrates. The minor V-shaped ZBA observed for the largest tip-sample separation in Fig. 10(b) also becomes wider as the tip-sample separation becomes smaller (also in the case of p-0.004 sample, data not shown), suggesting that this ZBA is (partly) a pseudogap feature, rather than (solely) a charging feature.

Contrary to the spectral features moving outward in Fig. 10(a), in smaller scale STS spectra shown in Fig. 10(b), the wavy features move to lower energy at smaller tip-sample separation. Such inward movement of the spectral features is consistent with the CS scenario [47]: at smaller tip-sample separation, the T-junction will have larger capacitance $C_T$, which reduces the peak separation of the different charge states in STS. The opposite behavior of the tip-sample separation dependent spectral features at different energy scales reveals that the ZBA on the p-0.008 and the p-0.004 substrate are a combination of pseudo-gap related features and a Coulomb charging effect. Due to the complicated residual Mott gap feature and the spectral broadening artifact induced by tip induce band bending, a quantitative modeling of the charging features is not possible at this time.

## VII    Conclusion

We have conducted a thorough STS investigation of hole-doped √3-Sn Mott insulating nanoscale domains on heavily-doped p-type Si(111) substrates. STS data at 77 K show that the hole-doping concentration is largest near the edges, as well as for the smallest nanoscale domains. STS spectra were recorded for different domain sizes, different doping levels, and different temperatures. For the √3-Sn surface on the most-heavily-doped substrate, small domains exhibit a strong ZBA, which evolves into a hard gap for the smallest 44 nm$^2$ domain. Two potential mechanisms for the ZBA are considered: the chiral $d + \mathrm{i}d$ wave superconductivity proposed in a recent theoretical work, and a DCB effect due to a charging effect in the √3-Sn domains. The latter results from the imperfect electrical coupling between the hole-doped Mott insulating domains and the Si substrate. Both scenarios fit well for weak ZBAs. However, the DCB theory provides a better fit when the ZBA becomes fully gapped. It also provides a better fit to the temperature-dependent tunneling data. A qualitatively different ZBA develops for √3-Sn domains grown on lesser-doped Si substrates (e.g. the p-0.004 and p-0.008 substrates). These spectra are interpreted in terms of a classical Coulomb blockade effect, where the resistances of the T-junction and S-junction are comparable.

With these observations, we conclude that the ZBA on the on the most-heavily-doped B√3 substrate is predominantly a DCB effect. Whether a secondary d-wave superconductivity-induced spectral feature exists in these spectra remains an open question. It is possible that superconductivity does not yet appear at this doping level or at this temperature. Comparing STS with and without a sufficiently high magnetic field is likely to clarify this issue. Alternatively, one could potentially use quasiparticle interference imaging to explore sign changes in the superconducting order parameter [48-50]. While sufficient small sizes are always expected to suppress Cooper pairing, superconductivity may ultimately emerge if one could increase the hole-doping level of large-scale domains well beyond the currently achievable maximum of about 12%. Such efforts are currently underway in our laboratory.

## ACKNOWLEDGMENT

This work was funded by the National Science Foundation under grant DMR 1410265.

FIGURES:

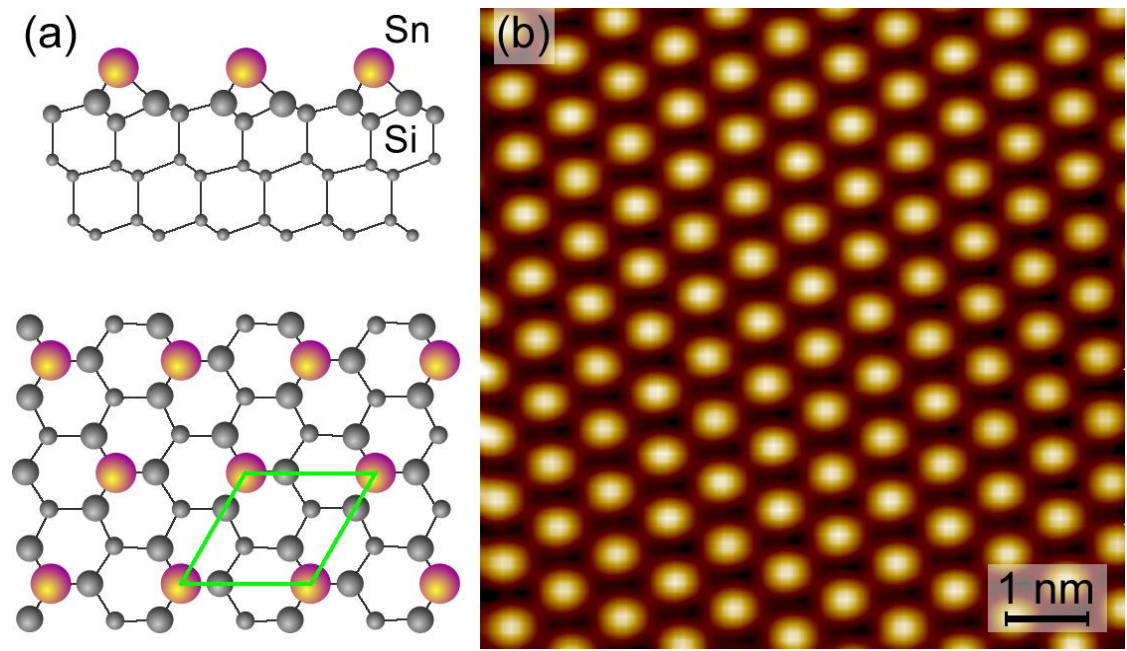

FIG. 1 (a) Atomic model of the √3-Sn structure. A unit cell is labeled. (b) STM image of the √3-Sn structure.

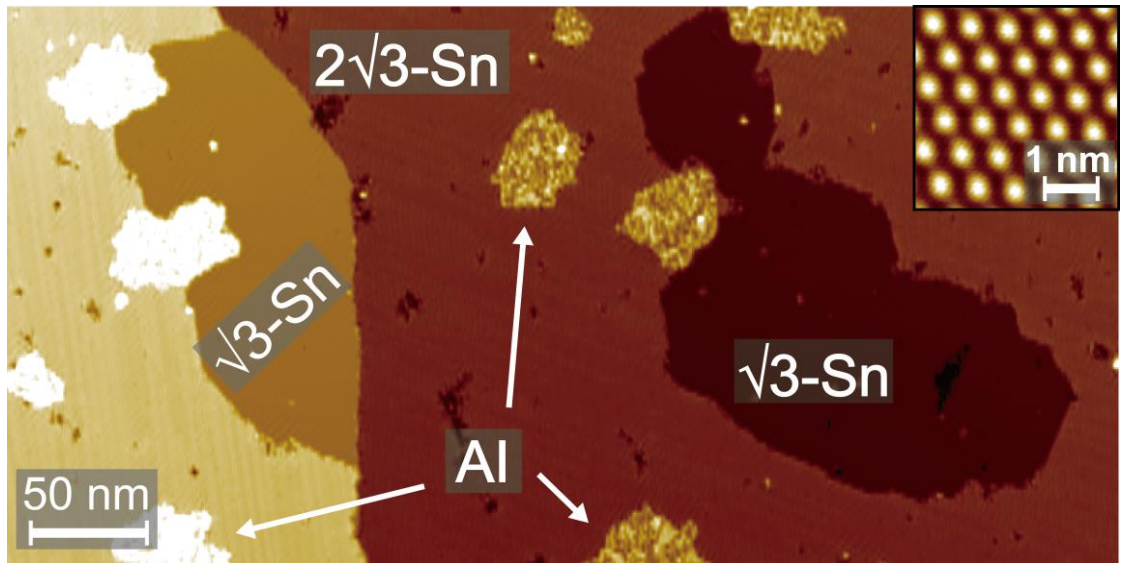

FIG. 2. STM image of a monatomic Sn layer on the B√3 substrate. The image contains a step edge left of center. The terraces mostly consist of the semiconducting 2√3-Sn phase decorated with disordered islands on top (labeled as "AI"). The √3-Sn nano-scale domains form isolated patches near the step edge (left), or inside the 2√3-Sn domains (right). The inset shows an atomic-resolution image of the √3-Sn structure.

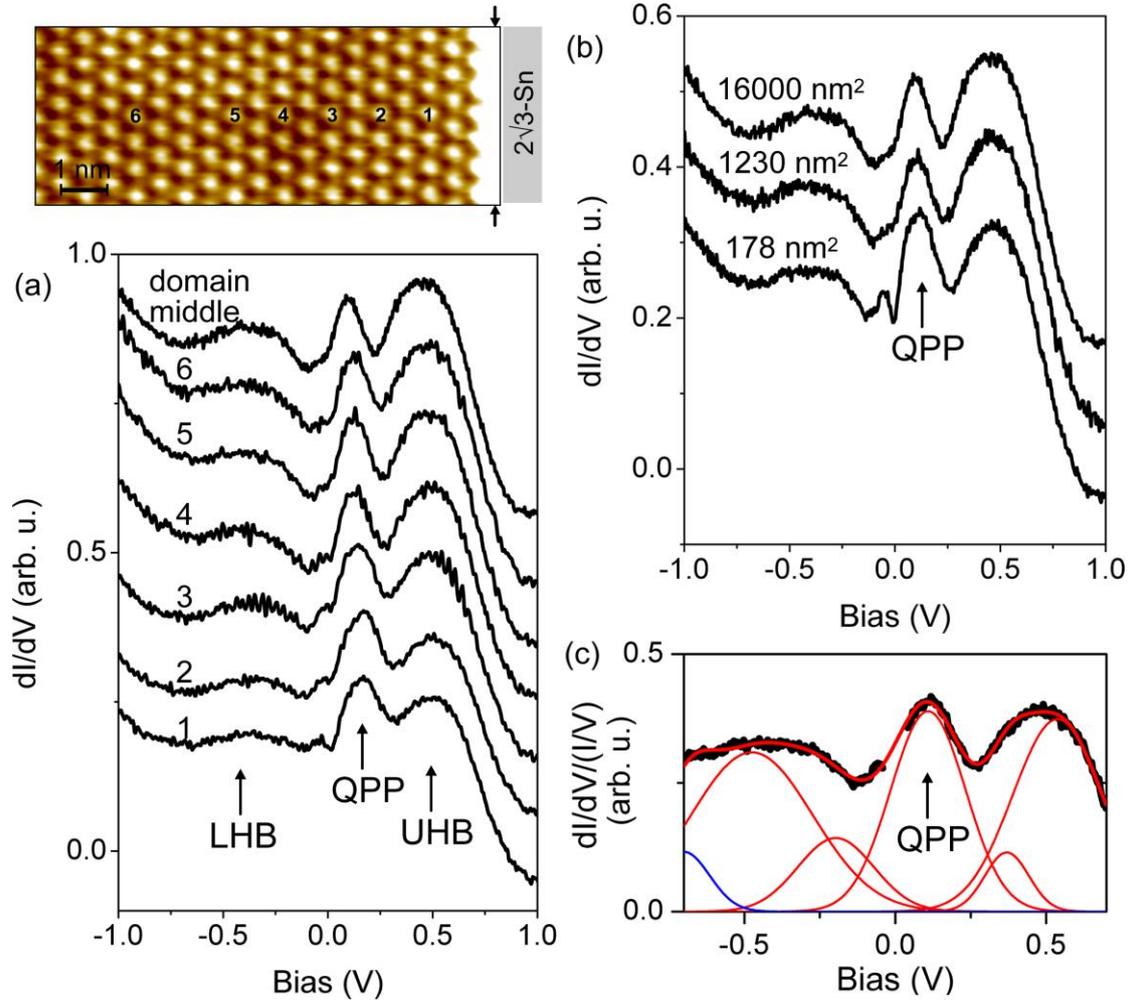

FIG. 3. STS characterization of the quasi-particle peak (QPP) and the two Hubbard bands at 77 K. (a) dI/dV spectra near the edge of a 16000 nm² √3-Sn domain, and bordering a 2√3-Sn domain. The domain edge is marked by arrows. The top curve is obtained near the middle of the domain which is out of the range of the STM image. The enhancement of the QPP in the lower dI/dV spectra is apparent from the peak height increase relative to the peak height of the UHB. (b) dI/dV spectra taken from the center of three √3-Sn domains, showing the enhancement of the QPP for the smallest domain. (c) Fitting of the 178 nm² spectra in (b) using six Gaussian peaks [21]. The dip feature between -0.04 V and + 0.04 V is excluded from the fitting. This dip is the ZBA feature, as discussed in the later part of the paper. Its influence on the spectral weight of the QPP is negligible. The single Gaussian peak for the QPP is labeled. The blue tail on the very left accounts for the bulk valence band. The other two Gaussians below the Fermi level account for the LHB while those to the right of the QPP account for the UHB. Curves in (a) and (b) are normalized to their intensity at ~ +1.5 V (outside the plot range), and are shifted vertically for clarity.

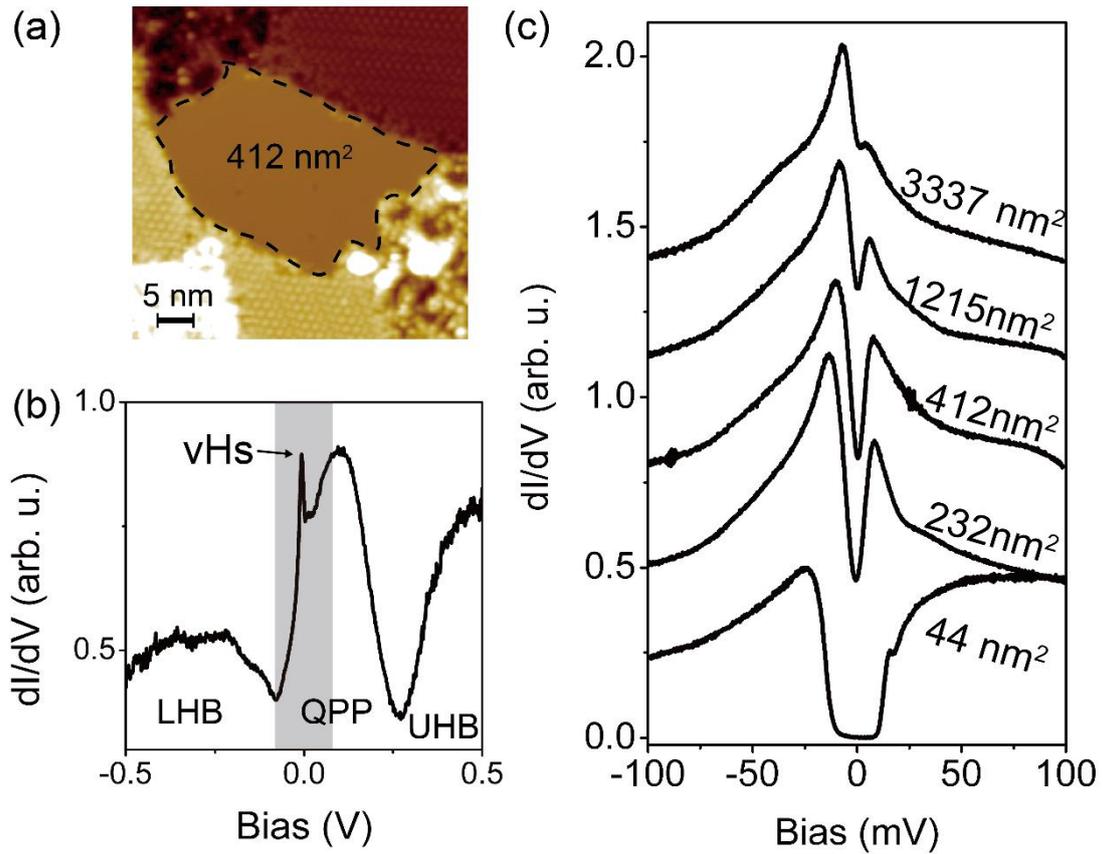

FIG. 4. STS measurements from different √3-Sn nano-domains on the B-√3 substrate at 4.4 K. (a) An STM image of a relatively small √3-Sn domain near a step edge. (b) Wide range dI/dV spectra showing the LHB, UHB and QPP. The sharp spike riding on top of the QPP is the van Hove singularity (vHs). The bias range displayed in (c) is marked in gray. (c) Small scale dI/dV data measured on different sized √3-Sn domains. The curves are normalized at their intensity near -100 mV, and are shifted vertically for clarity.

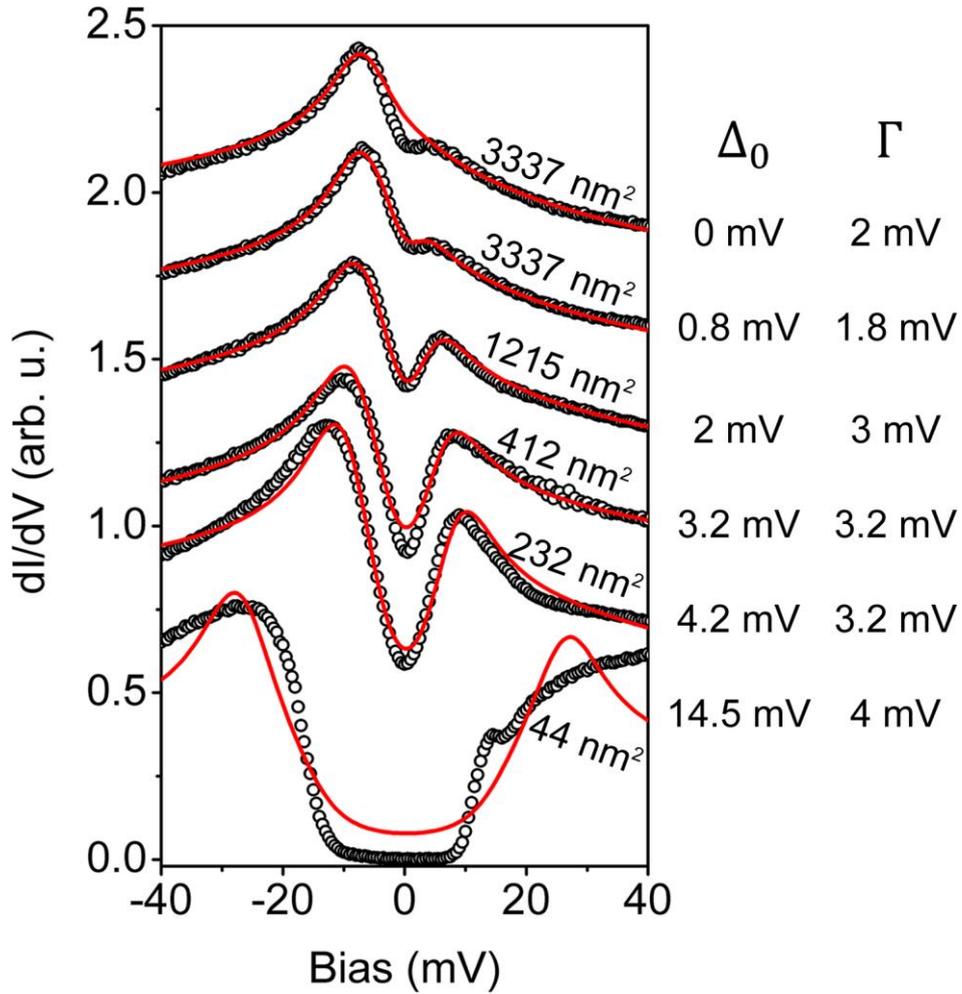

FIG. 5. Fitting of the 4.4 K STS data in Fig. 4(c) with the $d + id$ superconducting gap function (Eq. 2-4). The simulated spectra (red) are superposed onto the experimental curves (black circles). The best fitting parameters are indicated. The topmost fit to the 3337 nm² spectrum is conducted with fixed $\Delta_0 = 0$.

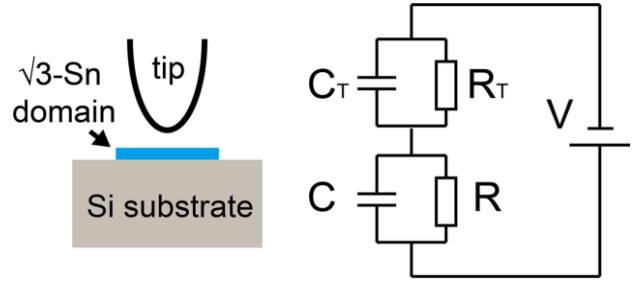

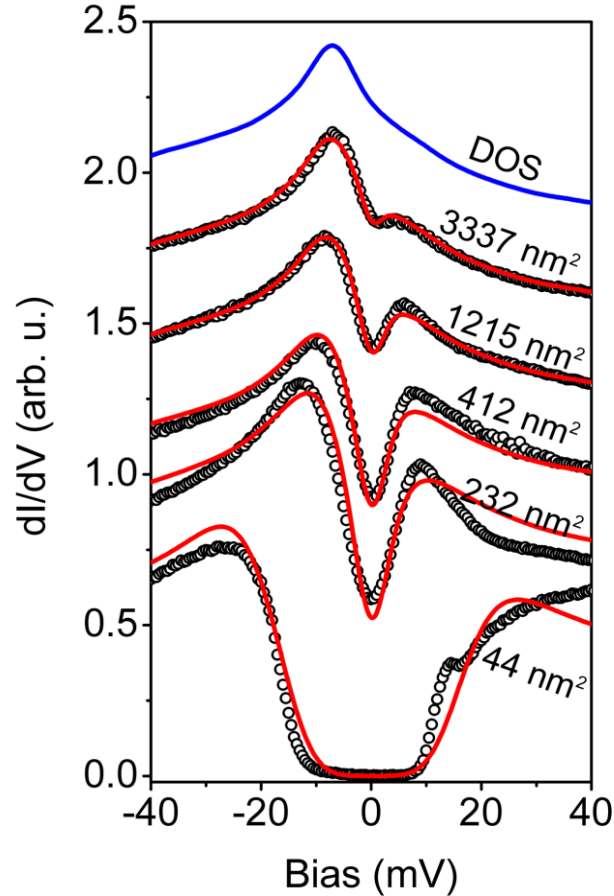

FIG. 6. Fitting of the 4.4 K STS data in Fig. 4(c) with the DCB model. The tunneling process is modeled by a double junction system, corresponding to the tip-√3-Sn domain junction (T-junction) and the √3-Sn domain-Si bulk junction (S-junction), as sketched on top. The lower panel shows the fittings (red) overlapping the data (black circle). The top curve is the DOS used for the fitting. The fitting parameters are: $T$ = 4.4 K; $C$ = 180 aF, $R$ = 0.06 $R_Q$ (3337 nm$^2$); $C$ = 72.5 aF, $R$ = 0.22 $R_Q$ (1215 nm$^2$); $C$ = 32.5 aF, $R$ = 0.56 $R_Q$ (412 nm$^2$); $C$ = 22.5 aF, $R$ = 0.96 $R_Q$ (232 nm$^2$); $C$ = 4.5 aF, $R$ = 4.9 $R_Q$ (44 nm$^2$). The curves are shifted vertically for clarity.

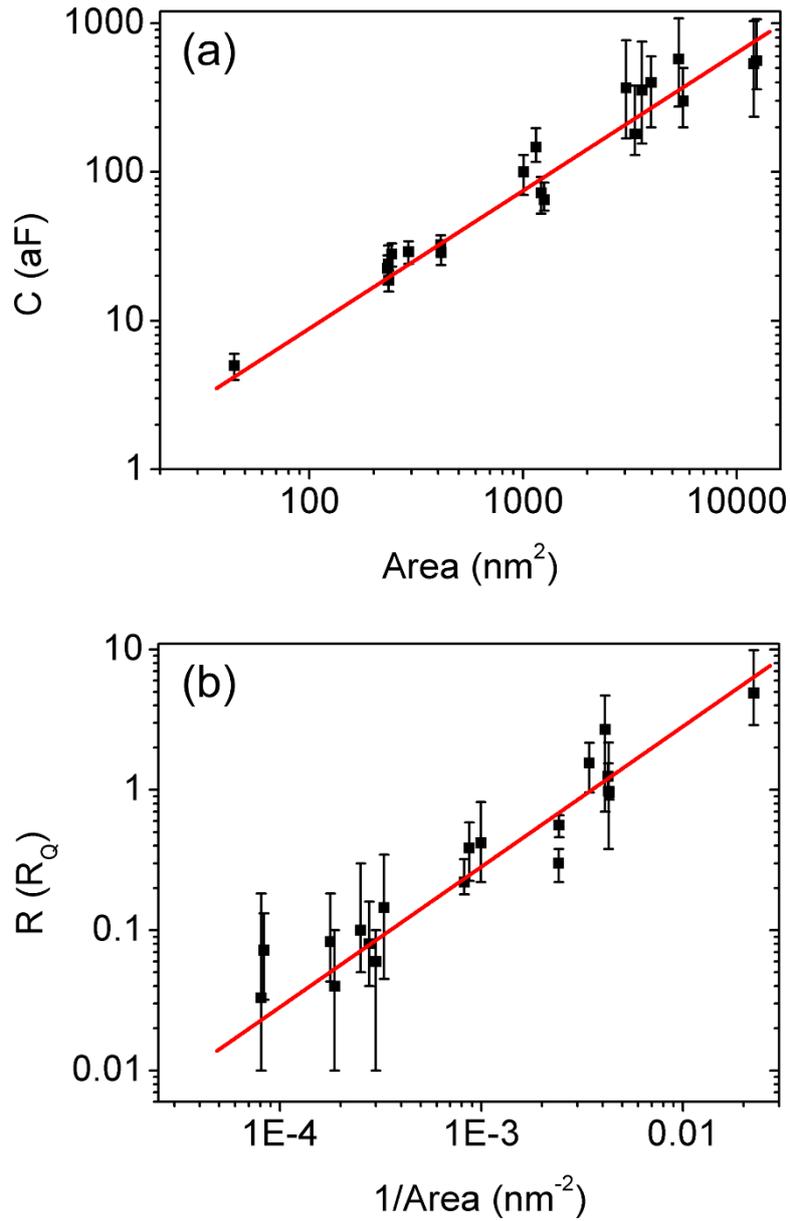

FIG. 7. Plot of the parameters obtained from fitting a series of STS spectra to the DCB model as a function of the √3-Sn domain size. Each data point represents the average of the fitting parameters, $C$ (a) and $R$ (b), estimated from multiple spectra taken from the same domain. The error bars represent the range of the best fit parameters. The red lines represent linear fits to the data on the double logarithmic scale.

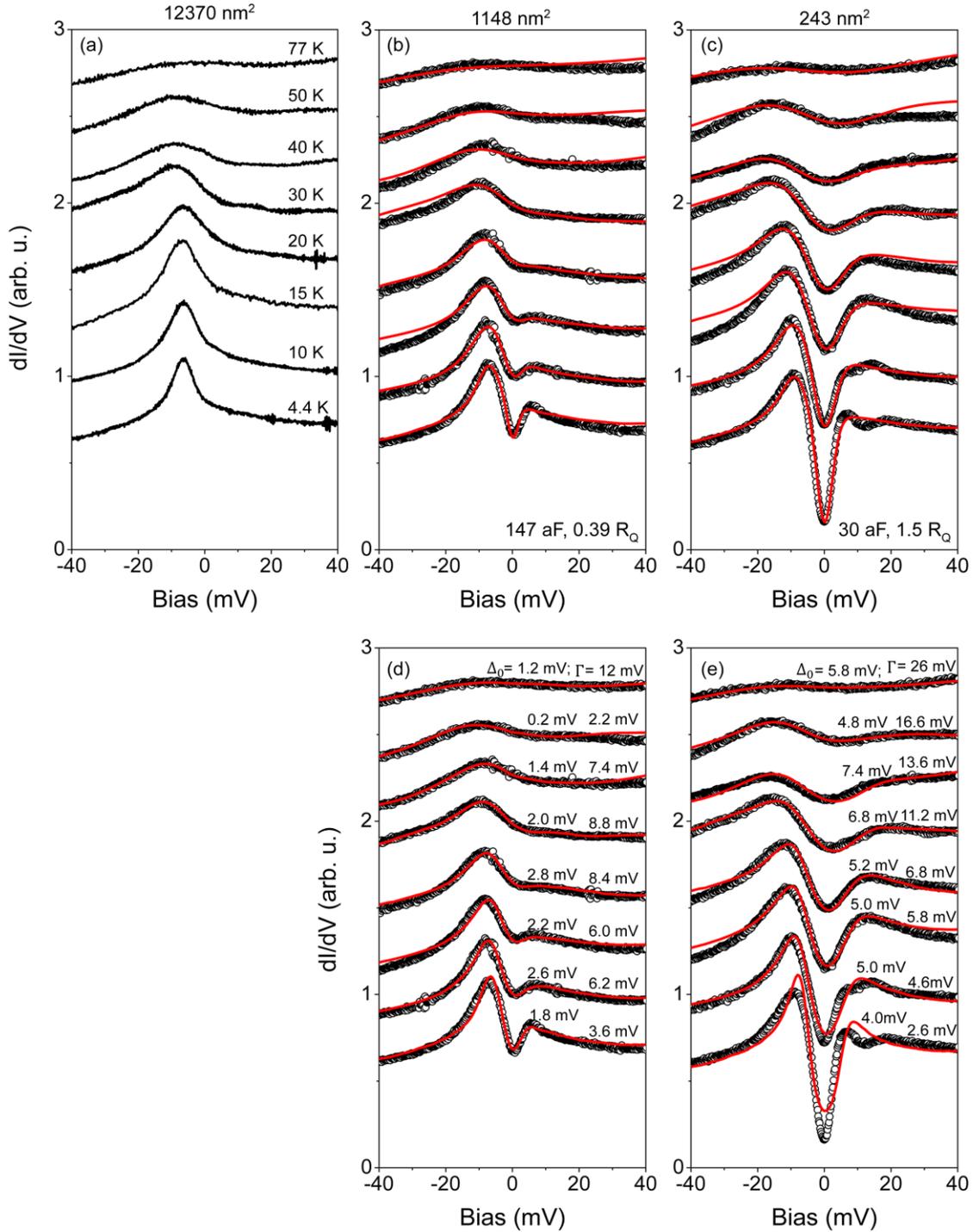

FIG. 8. Temperature dependent dI/dV spectra of three √3-Sn domains, with sizes of 12370 nm$^2$ (a), 1148 nm$^2$ (b,d) and 243 nm$^2$ (c,e). Temperatures indicated in (a) also apply to the other data sets following the same sequence. (b,c) Experimental data (black circles) on the two smaller domains are fitted (red) to the DCB model. The simulated spectra in each data set are computed using the capacitance and resistance values obtained from the DCB fit to the 4.4 K data. (d,e) Fitting the same set of data as in (b) and (c) but using the Dynes function (Eq. 11, 12). The resulting fit parameters are indicated.

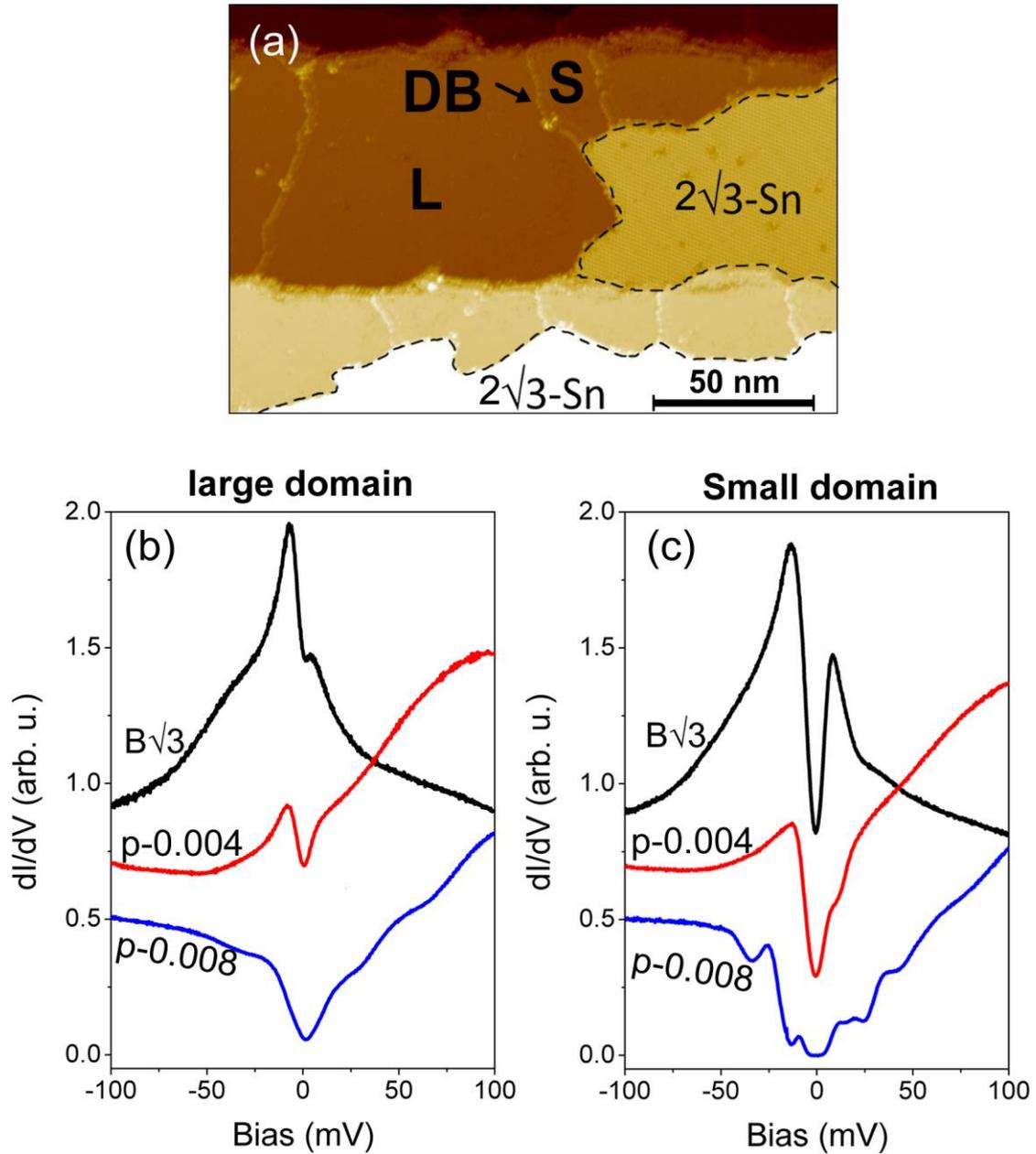

FIG. 9. Comparison of the dI/dV data from √3-Sn domains grown on p-type Si substrates with three different doping levels. (a) An STM image of the √3-Sn domains grown on the p-0.008 Si substrate. The surface contains a mix of the √3-Sn and 2√3-Sn phases. The 2√3-Sn domains are indicated with dashed lines. On the same terraces, neighboring √3-Sn domains are separated by domain boundaries (labelled as "DB"), appearing as bright lines. Two neighboring √3-Sn domains are labelled "L" and "S", representing large and small domains. (b) STS measured on relatively large √3-Sn domains (3337 nm$^2$, 6764 nm$^2$, 6000 nm$^2$ for the B-√3, p-0.004, p-0.008 substrates respectively). (c) STS measured on relatively small √3-Sn domains (232 nm$^2$, 356 nm$^2$, 234 nm$^2$ for the B-√3, p-0.004, p-0.008 substrates respectively). The curves in (b & c) are shifted vertically for clarity.

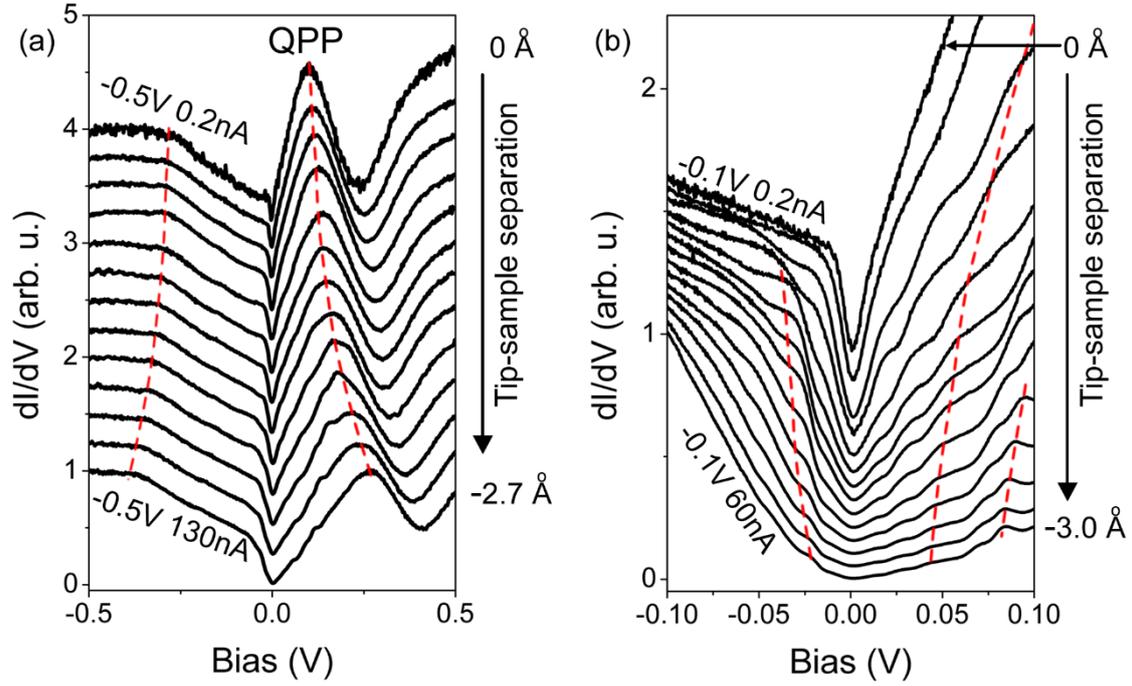

FIG. 10. Tip-sample-separation dependent STS measurements of a large √3-Sn domain (6000 nm$^2$) grown on the p-0.008 Si substrate, recorded at 4.4 K. (a) With decreasing tip-sample separation (or increasing set point current), the spectral features (including the QPP) shift to higher energy, implying a tip induced band bending effect in the Si substrate. (b) At smaller energy scale, the fine spectral features shift inward with decreasing tip-sample separation, implying a change in T-junction capacitance, $C_T$. In each plot, the curves are shifted vertically for clarity; the shifting features are tracked with dashed lines. Set point voltages and currents are indicated for the farthest and closest tip-sample distance. The tip-sample separation is also indicated on the right and is measured relative to the farthest tunneling set point.